\documentclass{raa}

\usepackage{graphicx,times}             
\input{epsf.sty}                        

\begin{document}

   \title{The site conditions of the Guo Shou Jing Telescope
}

   \volnopage{Vol.0 (200x) No.0, 000--000}      
   \setcounter{page}{1}          

   \author{Song Yao\inst{1,2}
   \and Chao Liu\footnote{Corresponding author}\inst{2}
   \and  Haotong Zhang\inst{2}
      \and Licai Deng\inst{2}
      \and Heidi Jo Newberg\inst{3}
      \and Yueyang Zhang\inst{1,2}
      \and Jing Li\inst{2,4}
      \and Nian Liu\inst{2,4}
      \and Xu Zhou\inst{2}
      \and Jeffrey L. Carlin\inst{3}
      \and Li Chen\inst{5} 
      \and Norbert Christlieb\inst{6}
      \and Shuang Gao\inst{2}
      \and Zhanwen Han \inst{7} 
      \and Jinliang Hou\inst{5}
      \and Hsu-Tai Lee\inst{8}
      \and Xiaowei Liu\inst{9}
      \and Kaike Pan\inst{10}  	
      \and Hongchi Wang \inst{11}
      \and Yan Xu\inst{2} 
      \and Fan Yang\inst{1,2}
      }

   \institute{Graduate University of Chinese Academy of Sciences, Beijing 100049, China\\ \and
     Key Lab of Optical Astronomy, National Astronomical Observatories, CAS, 20A Datun Road, Beijing 100012, China \\ \and
Department of Physics, Applied Physics, and Astronomy, Rensselaer Polytechnic Institute, 110 8th Street, Troy, NY 12180, USA\\ \and
	School of Physics and Electronic information, China West Normal University, 1 ShiDa Road, Nanchong, Sichuan 637000, China\\ \and
	Shanghai Astronomical Observatory, Chinese Academy of Sciences, 80 Nandan Road, Shanghai 200030, China\\ \and
	Center for Astronomy, University of Heidelberg, Landessternwarte, K\"onigstuhl 12, D-69117 Heidelberg, Germany\\ \and
	Yunnan Astronomical Observatory, Chinese Academy of Sciences, Kunming 650011, China\\ \and
	Academia Sinica Institute of Astronomy and Astrophysics, Taipei, China\\	\and
	Department of Astronomy \& Kavli Institute of Astronomy and Astrophysics, Peking University, Beijing 100871, China\\ \and
	Apache Point Observatory, PO Box 59, Sunspot, NM 88349, USA \\ \and	
	Purple Mountain Observatory, Chinese Academy of Sciences, Nanjing, Jiangsu 210008, China\\}

   \date{Received~~2012 month day; accepted~~2012~~month day}

\abstract{
The weather at Xinglong Observing Station, where the Guo Shou Jing Telescope (GSJT) is located, 
is strongly affected by the monsoon climate in north-east China. The
LAMOST survey strategy is constrained by these weather patterns.  In
this paper, we present a statistics on observing hours from 2004 to 2007, and
the sky brightness, seeing, and sky transparency from 1995 to 2011 at the site. We
investigate effects of the site conditions on the survey plan.  Operable
hours each month shows strong correlation with season: on average there are 8 operable
hours 
per night available in December, but only 1-2 hours in July and
August.  The seeing and the sky transparency also vary with seasons.  Although the
seeing is worse in windy winters, and the atmospheric extinction
is worse in the spring and summer, the site is adequate for the proposed scientific program of LAMOST survey.
With a Monte Carlo simulation using historical data on the site condition, we find
that the available observation hours constrain the survey footprint
from 22$^{\rm h}$ to 16$^{\rm h}$ in right ascension; the sky
brightness allows LAMOST to obtain the limit magnitude of
$V=19.5$\,mag with $S/N=10$.} 

   \authorrunning{S. Yao et al. }  
   \titlerunning{Guo Shou Jing Telescope Site Quality} 

   \maketitle

%
%
\section{Introduction}           
\label{sect:intro}

The Guo Shou Jing Telescope (also named as Large Sky Area Multi-Object
Fiber Spectroscopic Telescope, or LAMOST) is a quasi-meridian
reflecting Schmidt telescope with a 4-meter aperture, a 5-degree field
of view, and 4000 fibers installed on the focal plane
(\cite{cui10}; see also the overview by Zhao et al. 2012). It
will be used as a spectroscopic survey telescope for Galactic (Deng et
al. 2012) and cosmological science.

LAMOST is located at Xinglong Observing Station (here after XOS) of National Astronomical Observatories, 
Chinese
Academy of Sciences (NAOC). It is located $\sim170$ km northeast of
Beijing with longitude of $\rm7^{h}50^{m}18^{s}$ east and latitude of
$\rm40^{\circ}23^{\prime}36^{\prime\prime}$ north. 
The elevation of the site is about $\sim900$m.

The site weather is extremely seasonal and is dominated by a
continental monsoon in the summer. In this paper we show how the site climate, weather, sky brightness, and seeing influence the design of the LAMOST survey. These constraints are extremely important not only to the field selection for the halo on dark night (Yang et al. 2012) ,  on bright night (Zhang et al. 2012), and for the disk (Chen et al. 2012), but also for the science goals of the LAMOST survey (Deng et al. 2012). 



There have been a few previous papers on the XOS conditions. 
The early
work on monitoring the site conditions with a Schimdt telescope was
done by Zhou, Chen \& Jiang~(\cite{zhou00}). Liu et al.~(\cite{liu03})
used the the North pole monitoring data collected from 1995 to 2001 by
the 60/90cm Schmidt telescope on the same site to investigate the sky
brightness, the seeing and the atmosphere extinction. They found that
the mean sky brightness in $V$ band reaches 21.0\,mag arcsec$^{-2}$
and the mean seeing measured from the imaging FWHM varies with
seasons; specifically, the seeing is better in summers but worse in
winters. 
The camera used for seeing measurements on the Schmidt telescope
undersamples the point spread function, so that the seeing
measurements are not reliable when measured seeings are better than $\sim2$\arcsec.
The site seeing was also measured using a DIMM (Liu et al.~\cite{liu10}).

Previous studies either used old data or included only
a short time baseline. These measurements do not reflect the current weather
 and long term variation of the site conditions. Since LAMOST
will last at least 5 years, understanding the long term variation of
the site conditions is very important in designing and planning LAMOST
surveys.  In this paper we will revisit XOS quality as an astronomical
observing site using the largest and latest site parameter
data set available, covering almost 16 years since 1995.

In Section~2, we describe how the data is collected. In Section~3
the weather patterns, the sky brightness, the seeing, and the
atmospheric extinction are analyzed.
We then create a simplified but realistic simulation of the survey
with the priors of the site conditions so that the probability of the
sky coverage and the total number of the spectra are estimated, and
present that in Section~4. Finally, in section 5 we give
general guidelines that will be used in the design of the survey.

\section{Data Acquisition}

In this work we use two sets of data, both obtained from the BATC
survey (Beijing-Arizona-Taipei-Connecticut multi-color photometry
survey, Fan et al.~\cite{fan96}) using the Schmidt telescope at XOS. 

The observation logs of BATC from 2004 to 2007 provide the duration
of actual observing time from opening to closing the dome on each each night of operation. It is noted that the BATC is a 60/90\,cm Schmidt telescope on 
an equatorial mount, therefore it can point to whatever part of the sky
is clear, so that it is more time efficient than LAMOST.
LAMOST is limited to point only within 2 hours on each side of the meridian,
and is also restricted to -10$^\circ$ to 60$^\circ$ in
declination. 
(Higher declinations are possible, but at a significant cost to the field of view.)
Therefore, the actual observation time available for
LAMOST could be lower than BATC's statistics.

The other data set comes from Polaris area monitoring images collected by BATC
from 1995 to 2011. 
The BATC survey took an image of the north celestial pole before normal
observations on each night. Although it does not use the
zenith, which is a better direction for monitoring the sky
brightness, it is a great legacy to quantify XOS conditions.  Liu et
al.~(\cite{liu03}) used the first 6 years of data in their work. In the
current work, we extend it to 2011, 10 years more than was previously available.


We follow the same procedure used by Liu et. al.~(\cite{liu03}) to measure
the seeing, night sky brightness and extinction from the Polaris dataset:
  
 The sky brightness ADU was
measured from the background of the images, then instrument magnitude zero
points derived from photometric nights were interpolated to each observing
night to convert the sky brightness ADU into the magnitude in $V$ band.
The photometric software package SExtractor \cite{sextra} was used to measure the 
FWHM,  magnitude and other parameters of stellar objects on the image. 
The average FWHM of those stellar objects was taken as the seeing of that observation.
The sky transparency, which is quantified by the extinction
coefficient, was derived  by comparing  measured magnitudes of the selected stars with
the known magnitude of the same objects from Guide Star Catalog (GSC, Lasker et al. 2008).

\section{Site parameters}

\subsection{Weather patterns}
\label{subsect:weather}
\begin{figure*}[htbp]
\begin{center}
\includegraphics[scale=0.75]{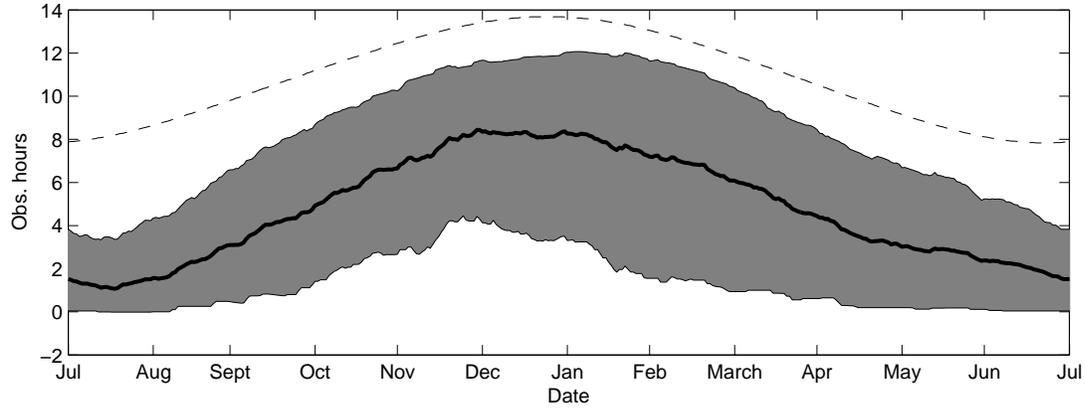}

\caption{
The statistics of the number of hours of BATC observations per night from
2004-01-01 to 2007-09-30. The thick solid line shows the actual mean
observing hours, smoothed with a 2-month window, at
each night from July 1 to Jun 30 the next year. The shaded area shows the
minimum and maximum hours of observation each night smoothed with the same time window. As a reference, the dashed line shows the theoretical
available time between evening and morning twilights. }\label{fig:obshours}

\end{center}
\end{figure*}

Fig.~\ref{fig:obshours} shows hours per night used for actual
observations by BATC from 2004 to 2007
Although the recorded
observed hours from BATC should not be identical to that for GSJT, it
does reflect the general weather pattern at XOS. 
Fig.~\ref{fig:obshours} shows that the actual observation
hours per night is about 6 hours less than the theoretical value 
at all times of the year. In this period of time, 43\% of observing time was lost to bad weather.
The maximum observing time is in
December and January, when about 8 hours on average are available each night,
while in July and August the average observing time is less
than 2 hours.  This pattern is similiar to
Figure~1 in Zhang  et al. (\cite{zhang09}) paper. 
Because it is restricted to observe within $\sim2$ hours of the meridian,
LAMOST hardly has any observation time in summer. Such a site dependent weather
pattern means that LAMOST will have a better chance to observe 
right ascensions that are observable in winter; the Galactic
anti-center and south Galactic cap. The north Galactic pole and the
inner Galactic disk, on the other hand, are difficult to completely sample with LAMOST.

\subsection{Sky brightness}
\label{subsect:sky}
\begin{figure*}[htbp]
\begin{center}
\includegraphics[scale=0.75]{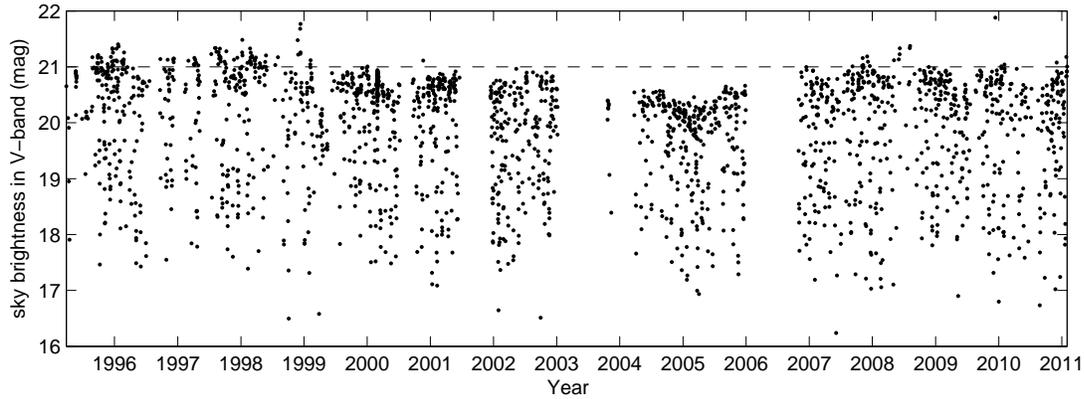}
\caption{The sky brightness in $V$ band obtained from BATC Polaris monitor data, as a function of time from 1995 to 2011. The dashed line is 21 mag arcsec$^{-2}$.}\label{fig:batcsky}
\end{center}
\end{figure*}

Fig.~\ref{fig:batcsky} shows the all sky brightness measurements from the BATC Polaris images as a function of
time. There are some long term features in the data. First, there is a
significant brightening between 2005 and 2006, which is due
to the construction of the buildings for LAMOST.
Lights from the construction site significantly affected the BATC telescope, which is only a few hundred
meters away. After the construction was done in 2007, the monitoring
data from BATC returned to the normal level for the site.

Second, there seems to be an annual (probably seasonal) variation in
the sky brightness. The periodical pattern is much clearer from 2007
to 2011. It seems that around the end of a year, i.e., in the winter,
the sky is darker, sometimes fainter than 21\,mag arcsec$^{-2}$, while in the
middle of a year, i.e., in the summer, is brighter by as much as 0.5\,mag.
It is not well understood what causes the seasonal pattern in the sky brightness.
Several factors could have contributed to the pattern, including dust storms in the spring, 
more construction at local areas in summer or local agricultural activities.
All of these factors reduce the transparency of the atmosphere, which
can result in more scattered light from nearby cities.

\begin{figure*}[htbp]
\begin{center}
\includegraphics[scale=0.75]{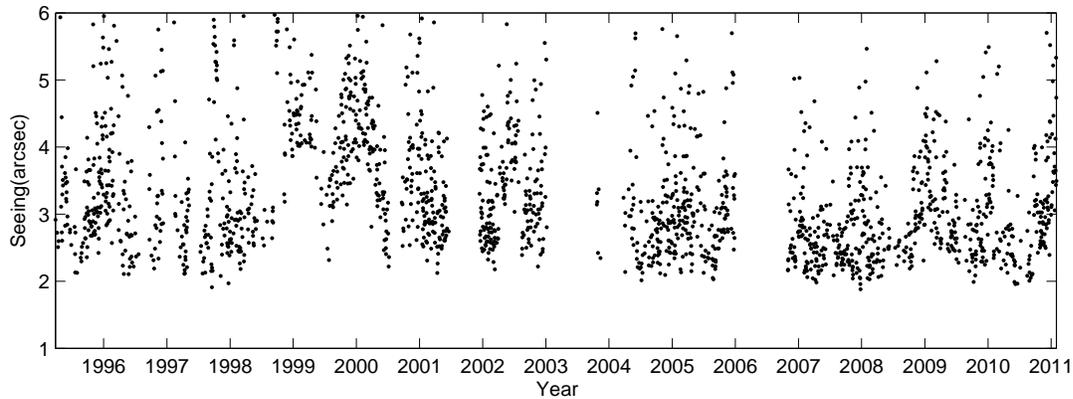}
\caption{The seeing measured from the FWHM of the Polaris-region images obtained by BATC from 1995 to 2011 is shown in this figure.}\label{fig:batcseeing}
\end{center}
\end{figure*}

\begin{figure}[htbp]
\begin{center}
\includegraphics[scale=0.75]{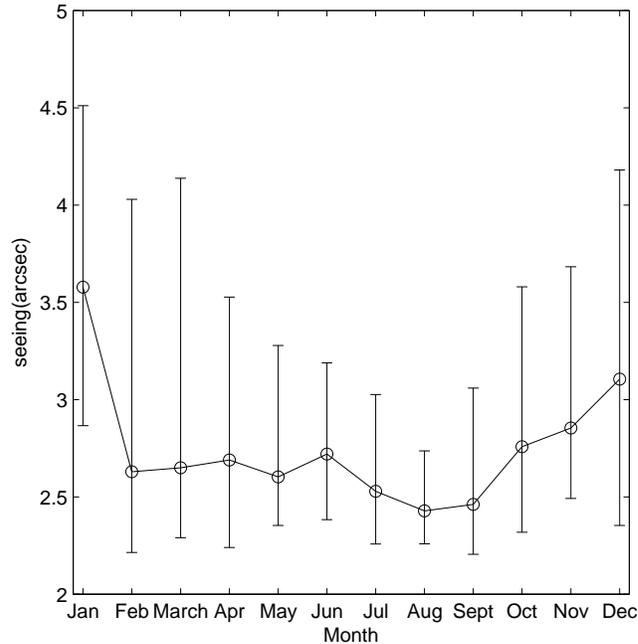}
\end{center}
\caption{The median seeing and its 1$-\sigma$ range using the data after 2007 in Fig.~\ref{fig:batcseeing} in each month are shown here.}\label{fig:batcseeingmonth}
\end{figure}

\begin{figure}[htbp]
\begin{center}
\includegraphics[scale=0.75]{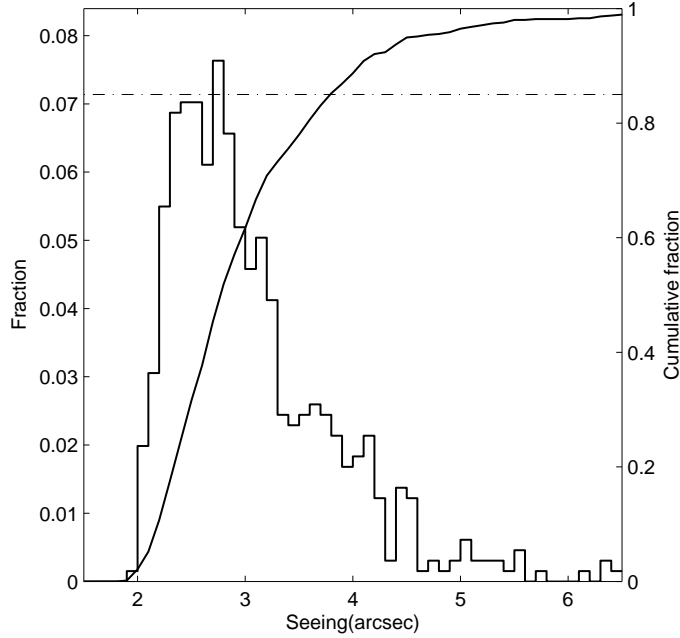}
\caption{
The distribution and cumulative distribution of the seeing after 2007,
using data from Fig.~\ref{fig:batcseeing}. The solid lines show the
distribution for all data. The dashed-dot horizontal line shows the
position of 85.3\%. It indicates that the 85.3\% of the observations
are in the seeing lower than 3.8\,arcsec for the
data.}\label{fig:batcseeinghist}
\end{center}
\end{figure}

\subsection{Seeing}
\label{sect:seeing}

There are no specific instruments dedicated to monitoring site quality, especially natural
seeing, at XOS. The most reliable data set that can be
used for seeing assessment is BATC data archive. 
However, note that the Schmidt telescope has a classical dome, with
attached office and living facility, so the measurements are likely
dominated by dome seeing.  The site natural seeing is most likely much  better than
 these measurements. The long term record of
seeing of XOS derived from BATC is shown in
Fig.~\ref{fig:batcseeing}. The seeing spans a huge range and is
probably seasonal, as is clearly seen in the figure. In the winters of
1999 and 2000, it was rather bad in general, and sometimes was worse
than 4\,arcsec. It became slightly better in 2006. There are several
possible explanations for the improvement. One possibility is that a 
new 4k$\times$4k CCD was
installed at the main focus the Schmidt telescope in that year.
The new camera improved the imaging resolution from 1.7\,arcsec/pixel to
1.36\,arcsec/pixel. This consequently slightly improves the
measurement of the seeing from the FWHM of the stars. An alternative
reason is that the weather since 2007 at XOS has been better.
Anyhow, in Fig.~\ref{fig:batcseeing} it is quite
clear that the data after 2007 have a slightly lower median seeing
than the earlier data, and also look more stable than before
2007. Therefore we only use the data after 2007 for subsequent analysis.

Fig.~\ref{fig:batcseeingmonth} shows the average seeing for each month
using the data after 2007. The best seeing occurs in August and
September, while the worst is in December and January. Meanwhile
the dispersion of the seeing (the error bar shown in the figure) also
follows the same trend. This pattern is similar to Fig~7 of Liu
et al.~(\cite{liu03}), and also Figure~5 of Zhang et
al.~(\cite{zhang09}). Experience from other telescopes at the same site
show that it's true that the seeing is worse in the winter and better
in the summer. This is related to the climate. In winter the
strong and frequent wind significantly enhances the turbulence in
the atmosphere and hence the seeing is worse and more
unstable. In contrast, in summers there is very weak or no wind and
this makes the seeing smaller and more stable. 
 
Fig.~\ref{fig:batcseeinghist} gives the statistics of the seeing after
2007. The peak of the histogram of the seeing is around
2.5\,arcsec. Based on the cumulative distribution of the seeing data,
85\% of the seeing measurements are better than 3.8\,arcsec.
Due to under-sampling, BATC images are not perfect for accurate seeing
measurements. Even with the slightly improved pixel scale of
1.36~arcsec, the images do not yield good measurements of the FWHM
below 2 pixels ($\sim$2.7 arcsec). Additionally, it is dome seeing
dominant and not the natural seeing at the site. For these reasons, we
will use the results only as a reference for the long-term variations
in seeing condition.Nevertheless, although the seeing is not perfect comparing with other famous sites, it is adequate for the proposed science program with LAMOST spectroscopic survey.

\subsection{Extinction}

\begin{figure}[htpb]
\begin{center}
\includegraphics[scale=0.75]{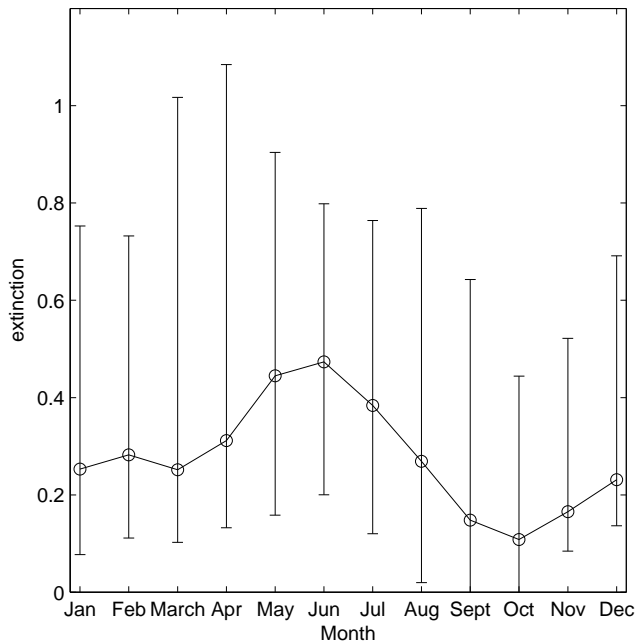}
\caption{The circles show the median atmospheric extinction coefficient for each month. The error bars show the range of 1-$\sigma$. Only the data after 2007 are used for this plot.}\label{fig:batcextinction}
\end{center}
\end{figure}




The sky transparency at XOS shows significant seasonal variations, 
as shown in Fig.~\ref{fig:batcextinction}. The best value occurs
in October, the worst value appears in May and June, when the local
area suffers the sand storms and/or climate factors such as humidity
and dust.





\section{A simulation for the LAMOST survey with site constraints}
\begin{figure}[htpb]
\begin{center}
\includegraphics[scale=0.75]{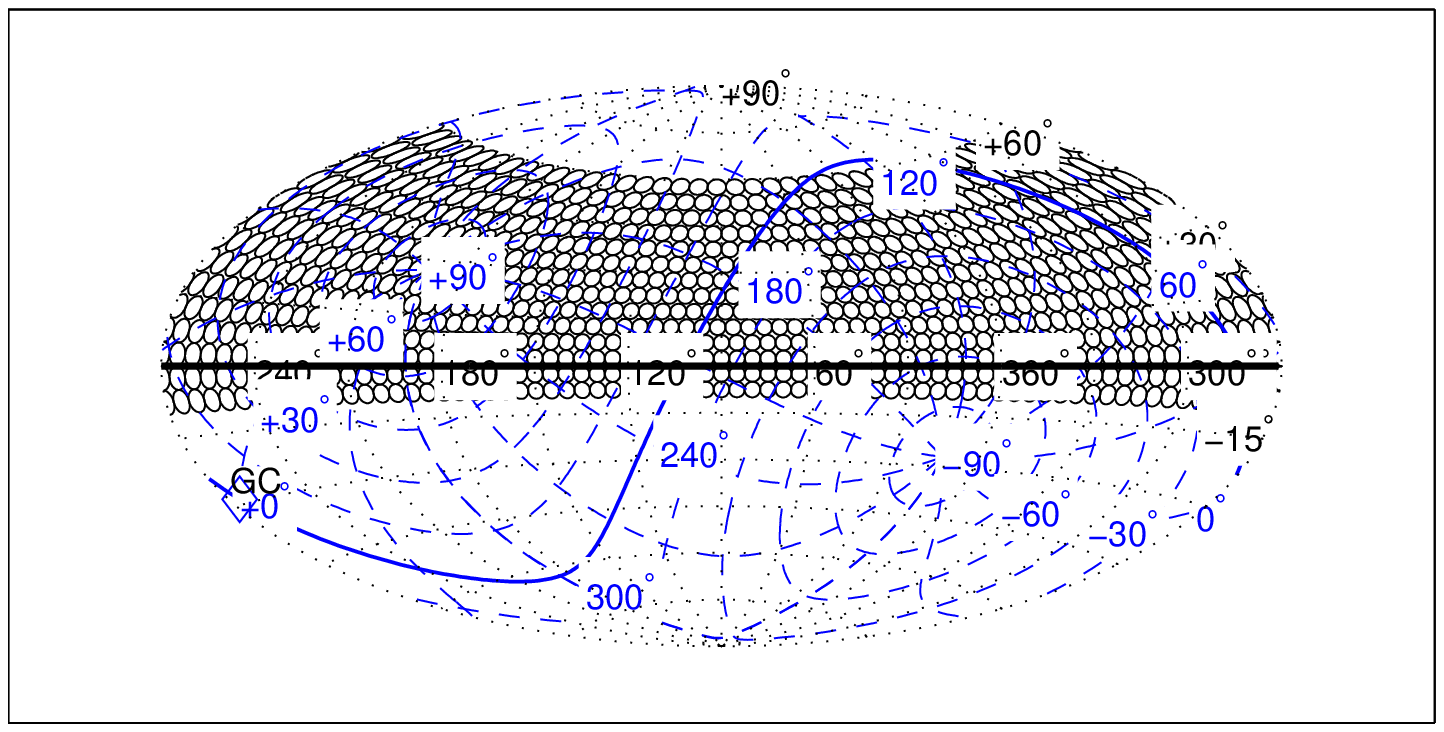}
\caption{The footprint map in equatorial aitoff projection of the Monte Carlo simulation. The circles show the possible plate that LAMOST can observe in the simulation. The blue grid shows the Galactic coordinates. Note that some plates are blocked by the coordinate labels when draw the plot, they do actually exist behind the numbers.}\label{fig:simfootprints}
\end{center}
\end{figure}

\begin{figure}[htpb]
\begin{center}
\includegraphics[scale=0.75]{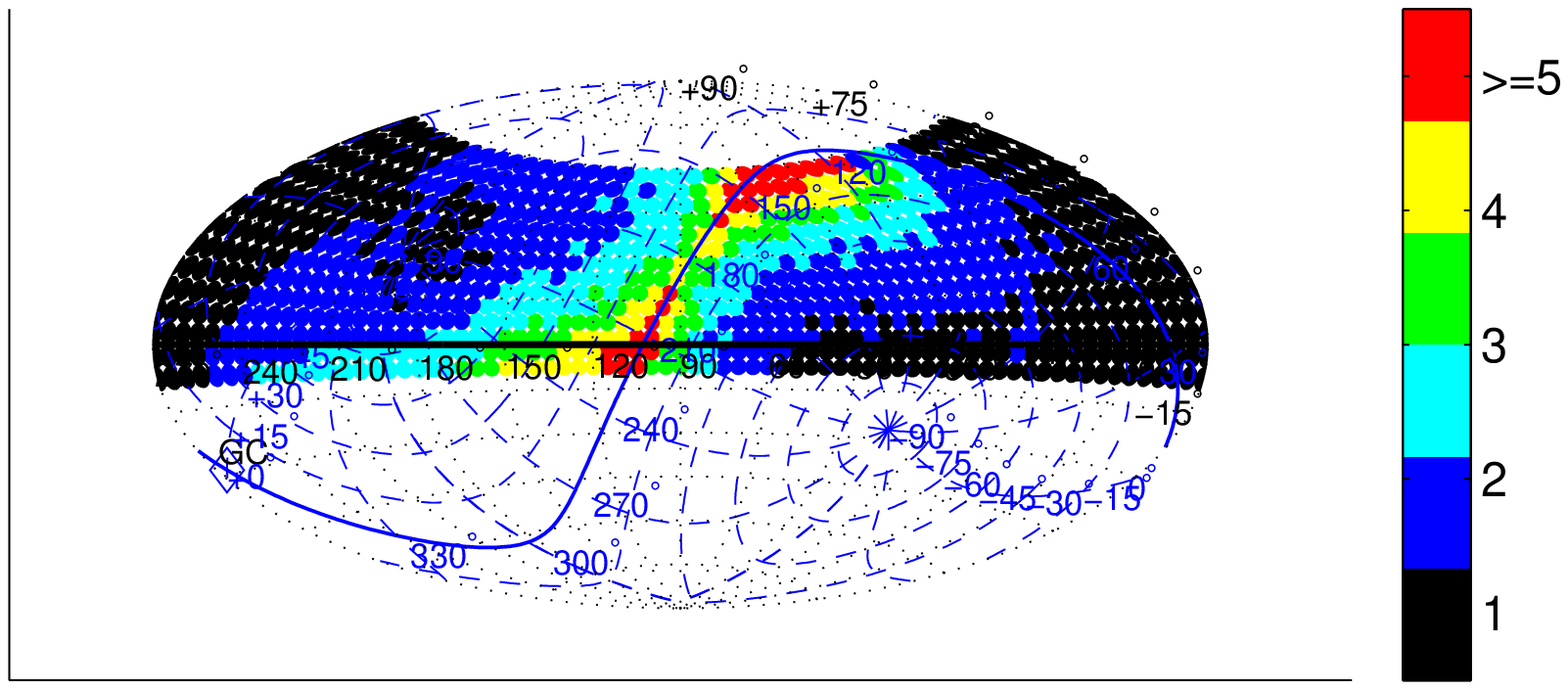}
\includegraphics[scale=0.75]{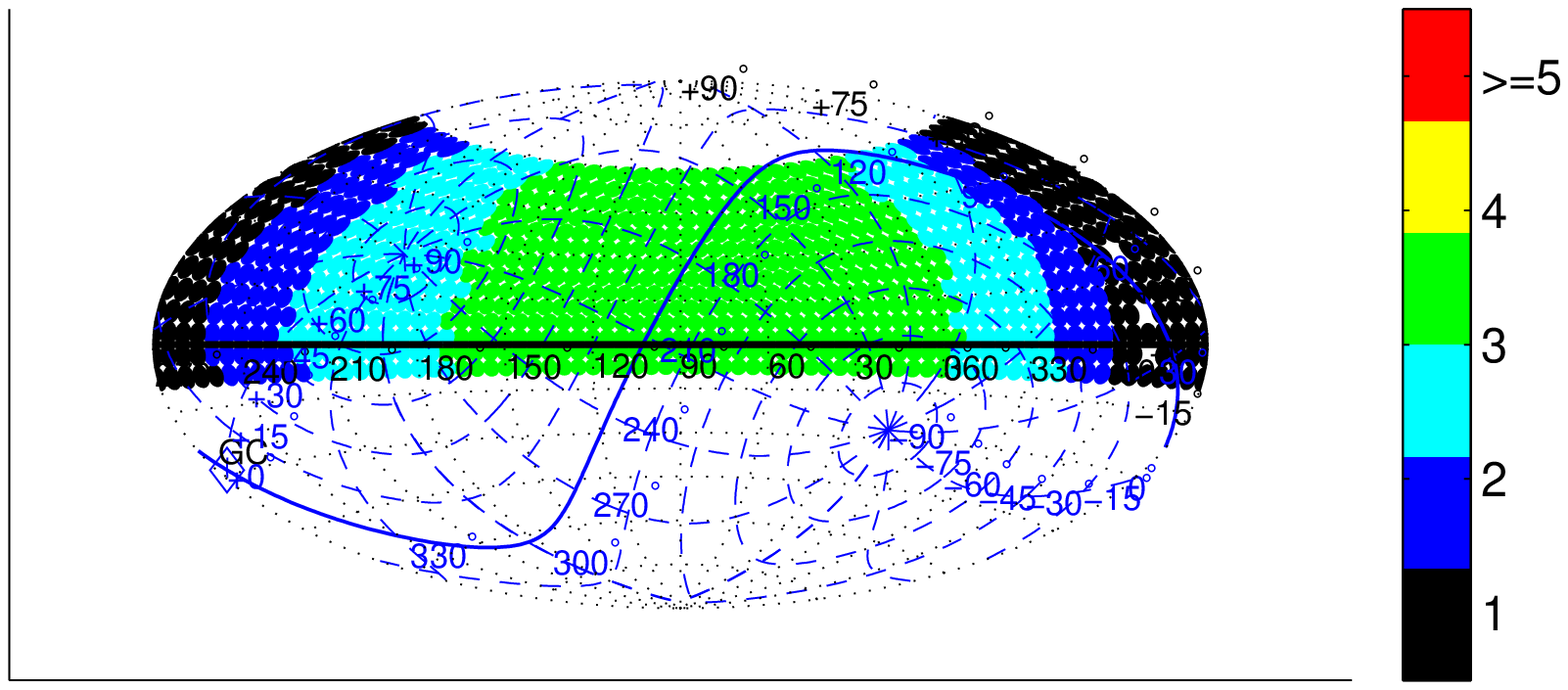}
\caption{\emph{Top panel: }The number density equatorial map of the observed plates at bright nights in five-year simulated survey. \emph{Bottom panel: }The number density equatorial map of the observed plates at dark/gray nights.}\label{fig:simnumberplates}
\end{center}
\end{figure}

We run a Monte Carlo simulation to simulate the sky coverage of the LAMOST survey based on the site conditions. The simulation helps to clarify how the site conditions affect the survey and consequently what science goals are the most feasible for LAMOST.


In the simulation, we assume that LAMOST can only observe 864
predefined non-overlapping 5-degree-diameter fields of view (hereafter denoted plates) covering all space between
$\delta=-10^\circ$ and $\delta=60^\circ$
(Fig.~\ref{fig:simfootprints}). This is an oversimplified assumption
since in practice LAMOST scans the sky with many overlapped
plates. However, this assumption makes the simulation quite simple and sufficient for investigating 
impacts of the site conditions on survey planning. No fiber assignment is included in the simulation. 
We also do not give any presumption of the favorite sky regions, e.g. the emphasizing of the anti-center 
direction as we did in Deng et al. (this volume),  but just evenly cover all available sky regions for LAMOST.  

We also assume that on dark/gray nights LAMOST observes the faint
objects with 1.5\,hour exposures (3$\times30$\,minutes), while at bright nights it only
observes the bright objects with 0.5\,hour exposure (3$\times10$\,minutes). The overhead for
each plate is 0.5\,hours, including telescope moving, active optics
operation, fiber positioning etc. In addition, we add 2 more hours for
each night for the telescope preparation and configuration,
including focusing the mirrors on a bright star twice per night.

We adopt a very simple strategy for the simulated survey. On dark/gray
nights, when LAMOST start to observe, it arbitrarily selects the one with the
minimum number of observations from all available plates within 2
hours on each side of the meridian. On bright nights, the plates with
zero observations will be the highest priority for observation.
Additionally, the probability of plate selection follows the star
counts in the area of sky covered
by that plate; therefore plates near the
Galactic mid-plane will be observed more than those in high latitudes.



We run the simulation for a five-year survey (2011--2015) and run it
50 times to smooth out the random fluctuations. The mean number of the
plates observed in five years, at each location, is shown in Fig.~\ref{fig:simnumberplates}. The
bright night survey can reach to more than 5 observations per plate
along the Galactic mid-plane. Since the Galactic plane, in particular
the anti-center direction, is right in the most weather favorable sky 
region for LAMOST, it seems that LAMOST is suitable for an anticenter Galactic
plane survey.
%
Since the dark/gray night survey does not use any prior knowledge of
the star distribution, it simply follows the 
actual number of observing hours available, considering the weather, seasonal variation in the number of hours per night, and moon phase.
In 5 years, the dark/gray night survey can cover
the sky between $\alpha=22^{\rm h}$ and $16^{\rm h}$ at least three
times.The rest of the region is the summer sky, which can be
essentially covered only once. 
The actual observation hours in the summer restrict the sky coverage of LAMOST.



\section{Conclusion}

From a compilation of site condition data in the last 16 years, we find that
the strongest constraint for the LAMOST survey is from the actual
observation hours, which is significantly lower than the available
hours due to weather constraints. This results in very few or even zero observation hours in
summer. Considering the special quasi-meridian design, the observable
sky is constrained to the autumn, winter and spring, so that
LAMOST performs best in the regions of the Galactic anti-center
 and the
South Galactic cap. Consequently, the survey plan has to be very
carefully designed under this limitation.

The sky brightness reaches roughly $V=21$\,mag arcsec$^{-2}$ on
the best nights at Xinglong. Therefore the limiting $V$ magnitude for spectroscopy should be
around 19.5\,mag with $S/N=10$ for point sources. This is the upper
limit of the capability of LAMOST.
Because the seeing at LAMOST site becomes worse during the months that
the actual observation hours are long, LAMOST will suffer from flux
lost due to the larger PSF, which will reduce the total throughput of
the fibers to some extent.

\begin{acknowledgements}
We thank the referee, Michael Ashley, for helpful comments on the manuscript.
This work is partially supported by CAS grant GJHZ200812,  Chinese National Natural Science
Foundation (NSFC) through grant No. 11243003, 10573022, 10973015,11061120454 and the US National Science foundation, through grant AST-09-37523.
\end{acknowledgements}

\appendix                  

\label{lastpage}

\end{document}